\documentclass[twocolumn,superscriptaddress,prl]{revtex4}
\usepackage{amssymb}
\usepackage{graphicx}
\usepackage{dcolumn}
\usepackage{bm}
\usepackage{amsmath}

\newcommand{\ket}[1]{| #1 \rangle}

\newcommand{\abs}[1]{\left|#1\right|}

\begin{document}

\title{Universal Spectra of Coherent Atoms in a Recurrent Random Walk}
\author{R. Pugatch}
\affiliation{Department of Physics of Complex Systems, Weizmann
Institute of Science, Rehovot 76100, Israel}
\author{O. Firstenberg}
\affiliation{Department of Physics, Technion-Israel Institute of
Technology, Haifa 32000, Israel}
\author{M. Shuker}
\affiliation{Department of Physics, Technion-Israel Institute of
Technology, Haifa 32000, Israel}
\author{N. Davidson}
\affiliation{Department of Physics of Complex Systems, Weizmann
Institute of Science, Rehovot 76100, Israel}

\begin{abstract}
The probability of a random walker to return to its starting point in dimensions one and two is unity, a theorem first proven by Polya \cite{Polya}. The recurrence probability --- the probability to be found at the origin at a time $t$, is a power law with a critical exponent $-d/2$ in dimensions $d=1,2$. We report an experiment that directly measures the Laplace transform of the recurrence probability in one dimension using Electromagnetically Induced Transparency (EIT) of coherent atoms diffusing in a vapor-cell filled with buffer gas. We find a regime where the limiting form of the complex EIT spectrum is universal and only depends on the effective dimensionality in which the random recurrence takes place. In an  effective one-dimensional diffusion setting, the measured spectrum exhibits power law dependence over two decades in the frequency domain with a critical exponent of $-0.56 \pm 0.01$. Possible extensions to more elaborate diffusion schemes are briefly discussed.
\end{abstract}

\maketitle

 A remarkable theorem due to Polya \cite{Polya} asserts that the probability of a random walker to return to the origin at dimensions $d=1,2$ equals one. Such walks are called recurrent walks. The Polya theorem has far reaching consequences in physics. For example, it is related to the well-known result that, in the non-interacting zero-spin Anderson model, disorder is strongly localizing at dimensions less than three \cite{PolyaLocalization1,PolyaLocalization2}. Strikingly, it can be shown that the absence of spontaneous symmetry-breaking phase-transition in $d<3$, the so called Hohenberg-Mermin-Wagner theorem, is also a consequence of Polya's theorem \cite{cassi}. Given the probability to be at the origin at a given time $t$, $P(\mathbf{r}=0,t)$, it can be shown that the walk is recurrent if and only if $\int_{0}^{\infty}P(0,t)dt$ diverges \cite{Redner}. Closely related to $P(0,t)$ are the distribution of First Passage Times, $\text{FPT}(\mathbf{r},t)$, which is the probability to reach the point $r$ for the first time at $t$, and the distribution of First Return Times, $\text{FRT}(t)=\text{FPT}(\mathbf{r}=0,t)$. These distributions are of great interest in many fields of research, from reaction-diffusion processes, the onset of firing bursts of neurons, porous media structure analysis, transport in disordered and scale invariant media, and the spreading of diseases \cite{Redner,KlafterNature}. In particular, FRT distribution determines trap performance in velocity-selective coherent population-trapping schemes \cite{LevyFlightVSCPT}.   

\begin{figure}[ht]
\begin{center}
\hspace{0.4cm}
\includegraphics[width=9cm, height=6 cm]{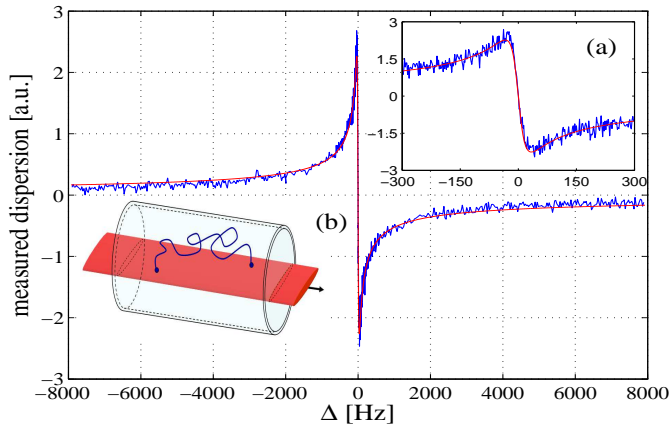} 
\end{center}
\par
\caption{Measured signal vs. the two-photon detuning $\Delta$ and the power law fit. Inset (a): zoom-in of the measured signal over the range $\pm 300$ Hz. Inset (b): an illustration of the one-dimensional beam experiment. Returning atoms contribute to the total intensity irrespective of the location of their reentry point thus effectively reducing the dimensionality to one.}
\label{fig:one}
\end{figure}

 Here, we report an experiment that directly measures the Laplace transform of the one-dimensional recurrence probability of a Rb atom diffusing in $\text{N}_2$ buffer gas. This is achieved by measuring the spectrum of Electromagnetically Induced Transparency (EIT) of a narrow light sheet (Fig. \ref{fig:one}). The complex spectrum thus obtained has a universal power-law form, which is the Laplace transform of the recurrence probability, $P(0,s) \propto s^{-\beta}, \beta = 0.5$, where $s=i\Delta+\Gamma_0/2$ is the complex Raman-detuning, with $\Delta$ being the two-photon Raman-detuning, and $\Gamma_0$ the EIT natural line-width. The measurements in the universal regime, where the power broadening and the beam size are small, depends only on $d$ --- the effective dimensionality of space. Our measurement yield this power law dependence with a critical exponent of $\beta=0.56 \pm 0.01$ over two decades in the frequency domain. In a control experiment, using a wide-area circular beam, the spectrum is expected to have power-law dependence of $s^{-\beta'}, \beta'=1$, which corresponds to a Lorentzian spectrum. In this experiment, our measurement yield $\beta'=0.97 \pm 0.01$.  
An explanation of these results is given using a sum over histories approach, which can be generalized to other systems where the atomic motion is not restricted to normal diffusion and can be either of the anomalous diffusion type or ballistic (in chaotic, mixed, or regular systems). An alternative derivation is also given by directly solving the relevant diffusion equation for the atomic coherences \cite{OferPRA}. Our results are an extension of the pioneering work of Xiao \textit{et. al.} \cite{RamseyDiffusion}, who recognized the importance of coherent returns as a line-shape narrowing mechanism, albeit not in the universal regime.  
  
 The probability of a diffusing particle that starts at the origin at $t=0$ to be at position $\mathbf{r}$ at time $t$ is given by $P(\mathbf{r},t)=(4\pi D t)^{-d/2} e^{-r^2/4Dt}$. 
It follows that the recurrence probability is $P(\mathbf{r}=0,t)=(4\pi D t)^{-d/2}$, which is a power law with an exponent $-d/2$. It is well known that the first passage time distribution is related to $P(\mathbf{r},t)$ and $P(0,t)$ by \cite{Redner}:

\begin{equation}\label{eq:FPT}
P(\mathbf{r},t)=\delta_{\mathbf{r},0}\delta_{t,0} + \int_0^t dt' \text{FPT}(\mathbf{r},t')P(0,t-t').
\end{equation}
This equation can be simply interpreted by noting that the probability to reach a point $\mathbf{r}$ at time $t$ is the sum over all intermediate times $t'$ of the probabilities to reach $\mathbf{r}$ for the first time at $t'$ and then to return to this point after a time $t-t'$. The first term in Eq. (\ref{eq:FPT}) is a boundary term for $t=0$.
To calculate the FRT, we substitute $\mathbf{r}=0$ in Eq. (\ref{eq:FPT}), take the Laplace transform from both sides, and recover 
\begin{equation}
\text{FRT}_d(s)=1-\frac{1}{P_d(0,s)}, 
\label{eq:FRT}
\end{equation}
where $P_d(0,s)$ is the Laplace transform of the recurrence probability in dimension $d$. For $d=1$, $P_{1}(0,s) = (4Ds)^{-1/2}$ which is a power law in $s$, while for $d=2$ $P_2(0,s)=K_0(\sqrt{s/\Gamma_D}) \sim \text{ln}(s/\Gamma_D)/4-0.577$ for small $s$, where $K_0$ is the modified Bessel function of the second kind, $\Gamma_D=\text{4D}/\text{W}_0^2$ is the diffusion broadening, $D$ is the diffusion coefficient, and $W_0$ is the beam diameter. 
 
 In brief, our EIT is obtained with two phase-coherent laser fields that couple the two magnetically insensitive hyperfine ground-state levels of $^{87}\text{Rb}$: $\ket{1}=\ket{5^{2}S_{1/2},F=1,m=0}$ and $\ket{2}=\ket{5^{2}S_{1/2},F=2,m=0}$ to a mutual excited level $\ket{3}=\ket{5^{2}P_{1/2},F'=2,m=1}$. When the energy difference between the two beams is equal to the ground level splitting, the atom is pumped towards a dark state $\ket{D}=2^{-1/2}(\ket{1}-\ket{2})$ at a rate $\Gamma_{p}$, which is proportional to the laser intensity. An atom in a dark state decouples from the laser fields, and is therefore transparent. An orthogonal state to $\ket{D}$ is the bright state $\ket{B}=2^{-1/2}(\ket{1}+\ket{2})$, which can be excited to $\ket{3}$. An atom in a dark state that wanders out of the EIT beams will perform dark-to-bright oscillations at a rate determined by the two-photon Raman-detuning, $\Delta$, in the rotating frame defined by the two lasers. 
 
  An atom in the dark state that stays within the beams for a time $t$ has a probability $\text{Re}(e^{-s_\text{in}t})$ to be in a dark state upon leaving, where $s_\text{in}=-i\Delta+(\Gamma_0+\Gamma_{p})/2$. An atom in a dark state that leaves the EIT beams for exactly $T$ seconds has a probability $\text{Re}(e^{-sT})$ to be in a dark state upon returning, where $s=-i\Delta+\Gamma_0/2$ \cite{RamseyCPT}.  
If the pumping rate back to dark state, $\Gamma_{p}$, upon returning to the beam, is much smaller than the dark-state decay rate, $\Gamma_0$, the total transparency is proportional to $\text{Re}(e^{-sT})$. 

 In our experiment, the EIT beams are localized to a small volume in space. The atoms diffuse in and out of the beam randomly due to the collisions with the buffer gas. The steady-state EIT spectrum is a sum over all possible histories of dark-state atoms being within the beam for \textsl{exactly} $t$ seconds, then staying outside the beam for \textsl{exactly} $T$ seconds, then coming back for exactly $t'$ second, staying outside for \textsl{exactly} $T'$ seconds, and so on. For example, the contribution to the spectrum of atoms that were pumped to dark state after passing through the beam and then lost their coherence outside the beam is $S(s)=\int_{0}^{\infty}dt\text{FPT}(t)e^{-s_\text{in}t}=\text{FPT}(s_\text{in})$, where $\text{FPT}(t)$ is the average FPT distribution through a beam of diameter $W_0$ \cite{remark}. 
Summing over all histories, we obtain
\begin{equation}
S(s) = \sum_{n=0}^{\infty} \text{FPT}(s_\text{in})^{n+1} \text{FRT}(s)^n = \frac{\text{FPT}(s_\text{in})}{1-\text{FPT}(s_\text{in}) \text{FRT}(s)}.
\label{eq:SumOverHistories}
\end{equation}
Note that long trajectories are less probable due to the decay factor.   
 
  In the limit when the beam size tends to infinity with uniform intensity, there are no dark periods and thus the FRTs drop from Eq. (\ref{eq:SumOverHistories}). The obtained complex spectrum is a power law, $S=\text{FPT}(s_\text{in})=s_\text{in}^{-1}$, which is the well-known complex Lorentzian spectrum, \textit{i.e.}, the absorptive part of $s_\text{in}^{-1}$ is the Lorentzian distribution. The width of the Lorentzian is $\Gamma_0+\Gamma_p$, where $\Gamma_p$ is recognized as the power broadening. 
  
  In contrast, the universal limit is obtained when the beam size tends to zero $\Gamma_D \gg \Gamma_0$, with a uniform and low intensity, such that $\Gamma_{p} \ll \Gamma_0$. In this limit, the transit-time broadening is large and $\text{FPT}(s_\text{in}) \rightarrow 1$. The dependence on the transit time drops from Eq. (\ref{eq:SumOverHistories}), and the spectrum becomes universal:

\begin{eqnarray}\label{eq:SumOverHistories2}
S(s) = \sum_n \text{FRT}(s)^n = \frac{1}{1- \text{FRT}(s)} = P(0,s). 
\end{eqnarray}

 We now turn to an alternative derivation of Eq. (\ref{eq:SumOverHistories2}). In \cite{TopologicalStability} we demonstrated that the evolution of the atomic coherence field during storage-of-light is governed by a diffusion equation. In \cite{OferPRA}, we developed a comprehensive theory for EIT spectra of thermal atoms. The resulting diffusion equation for the coherence and the complex spectrum that directly depends on it are given by

\begin{gather}
\left[ -i\Delta +\frac{\Gamma_0}{2} -D\nabla_{\bot}^{2}+\Gamma _{p}\left( \mathbf{r}\right) 
\right] R_{21}\left( \mathbf{r}\right) =-n_{0}\Gamma _{p}\left( \mathbf{r}
\right) ,  \label{eq_R21_2} \\ \nonumber
S(s) \propto \int d^{d}\mathbf{r~}\Gamma _{p}\left( \mathbf{r}
\right) \left[2R_{21}\left( \mathbf{r}\right) /n_{0}\right] ;
\label{EqFromPra} 
\end{gather}
where $\Gamma_p(\mathbf{r})$ is proportional to the intensity profile, $R_{21}\left(\mathbf{r}\right)$ is the steady-state ground state coherence, $n_0$ is the density ($\sim 9 \cdot 10^{10} / cc$), and $S(s)$ is the normalized complex spectrum. In the universal limit, we neglect $\Gamma_p(\mathbf{r})$ from the left hand side of Eq. (\ref{eq_R21_2}) and replace $\Gamma_p(\mathbf{r}) \rightarrow \Gamma_p \delta(r)$. We then obtain
\begin{equation}\label{eq:DiffusionThreeLevel2a}
[-i\Delta+\frac{\Gamma_0}{2}-D\nabla_{\bot}^2]R_{21}=-n_0 \Gamma_p \delta(r),
\end{equation}
Taking the Fourier transform, algebraically inverting the equation, and taking the inverse Fourier transform, we find
\begin{eqnarray}\label{eq:DiffusionThreeLevel2b}
R_{21}(r) = -n_0 \Gamma_p \int d^dq \frac{e^{-i\vec{q} \cdot \mathbf{r}}}{s+Dq^2},  \\ \nonumber
S(s) \propto  \int_{0}^{\infty} d^dr R_{21}(r) \delta(r) = P_{d}(0^{+},s).
\end{eqnarray}
This equation can be solved exactly in dimensions $d=1$ and $d=2$. In $d=1$, $S(s) \propto s^{-1/2}$ and for $d=2$, $S(s) \propto K_0(\sqrt{s/\Gamma_D}) \sim \text{ln}(s/\Gamma_D)/4-0.577$ for small $s$, in agreement with the sum-over-histories approach [Eq.(\ref{eq:SumOverHistories2})]. This agreement stems from the fact that one can always write the diffusion propagator as a path integral. Taking the beam size to infinity while keeping the intensity finite, we can neglect the diffusion by setting $D=0$ in Eq. (5). The resulting complex spectrum is the complex Lorentzian spectrum, $S(s) \propto s_\text{in}^{-1}$. 

\begin{figure}[h]
\begin{center}
\hspace{0.4cm}
\includegraphics[width=9.2cm, height=6.2 cm]{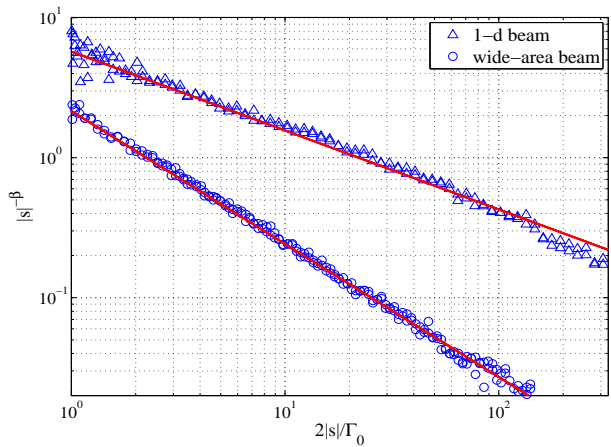} 
\end{center}
\par
\caption{Measured signal for the $1d$ beam configuration (triangles), and for the wide area beam (circles) vs. the complex detuning size $\abs{s}$ in units of $\Gamma_0/2$. The lines are theory with a single fit-parameter --- the critical exponent. We find $\beta=0.56 \pm 0.01$ over two decades in frequency and $\beta'=0.97 \pm 0.01$ over two decades.}
\label{fig:two}
\end{figure}

 Our experiment consists of a single mode Vertical Cavity Surface Emitting Laser (VCSEL) tuned to $\sim 795$ nm on the $^{87}\text{Rb}$ D1 transition. By modulating the VCSEL's current we created two equally powered sidebands at $ \sim \pm 3.4 $ GHz, which consist our co-propagating EIT modes (separated by the hyperfine splitting of $\sim 6.8$ GHz). A secondary weak frequency-modulation (FM) at $97$ kHz was added to facilitate low noise lock-in detection. Using weak ($1\%$) modulation depth ensured that the measured beat signal is directly proportional to the dispersive part of the EIT spectra \cite{BenAroya}. The beam was circularly polarized and passed through an isotopically pure $^{87}\text{Rb}$ vapor cell with $10$ Torr of $N_2$ buffer gas ($D=10$ $\text{cm}^2/\text{sec}$). The cell dimensions were $2.5$ cm in diameter and $5$ cm in length, water-heated to $\sim 48 C$, and magnetically isolated using three concentric $\mu-$metal shields. A set of Helmholtz coils were used to provide a homogeneous longitudinal magnetic field in the beam propagation direction of $49$ mG, thus pushing the magnetic sensitive EIT lines $\pm 69$ kHz sideways. The line-width broadening due to the inhomogeneous second order Zeeman shift is less than $1$ Hz, much smaller than the line-width, which is essential for observing the universal regime. After exiting the cell, the EIT beam was focused on a photo-detector (PD). The PD signal was filtered and amplified using on-board electronics and was fed to a dual-channel lock-in amplifier (SRS model SR-850) to allow for demodulation of the $97$ kHz EIT beat signal.
 
 In our measurements, we scanned the two-photon detuning $\Delta$ in the range $\pm 8$ kHz in frequency steps of $2-4$ Hz. The dwell time per detuning was $0.5$ sec. After a proper settling time of typically $10$ msec, we sampled and subsequently averaged the lock-in signal and obtained the dispersive part of the EIT spectrum as a function of $\Delta$. In the first experiment, we used a wide area beam with a diameter of $8$ mm. We measured the spectrum for different powers and found the power broadening slope to be $165$ Hz/(W/m$^2$) and $\Gamma_0 = 45$ Hz. In the second experiment using a spherical and a cylindrical lens, we shaped the beam to become a light sheet with beam diameters of $126$ $\mu m$ $\times 1$ cm along the cell length [Fig. \ref{fig:one}(b)]. Atoms returning to the beam contributes to the transparency irrespective to the location of their reentry point \cite{zmotion} making this geometry effectively one-dimensional.   
The diffusion broadening, determined by the narrow dimension of the beam $W_0=126 \mu$m, is $\Gamma_D=4D/W_0^2=252 \text{kHz} \gg \Gamma_0$ as required. 
The Rayleigh range of the beam is $\pm 6.3$ cm from the cell center, so the beam expands by less than $8\%$ within the cell, thus keeping a well-defined width along the cell. 
We adjusted the power of the light sheet to $150$ nW such that the power broadening is $\Gamma_p=20$ Hz. 

\begin{figure}[ht]
\begin{center}
\includegraphics[width=9cm, height=6cm]{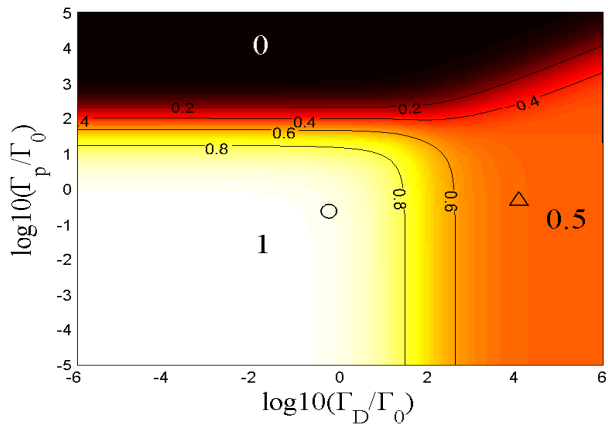}
\end{center}
\caption{Map of critical exponent $\beta$ calculated from the exact theory [Eq. (5)] as a function of the diffusion broadening $\Gamma_D$ and the power broadening $\Gamma_p$ in units of $\Gamma_0$. We extract $\beta$ from the calculated spectrum using linear fit, with $\Delta$ in the range $\pm 8$ kHz. 
The linearity of the fit significantly increases towards zero and infinity. When $\Gamma_{p} \rightarrow \infty$ at a fixed $\Gamma_D$, $\beta \rightarrow 0$, signifying flat spectrum over the fit range. When the beam size tends to infinity, ($\Gamma_{D} \rightarrow 0$) at a fixed $\Gamma_p$, the line-shape becomes a Lorentzian, and $\beta \rightarrow 1$. The universal regime is where 
$\beta \rightarrow 0.5$. The triangle (circle) indicates the position where we performed our $1-d$ (wide-area) measurements. The expected slope according to this prediction is $0.52$, compared to the measured $0.56 \pm 0.01$. A similar calculation for the wide area beam yields $0.99$, compared to the measured $0.97 \pm 0.01$.}
\label{fig:three}
\end{figure}

 In Fig. \ref{fig:one}, we plot the normalized measured dispersive spectrum of the $1d$ beam configuration as a function of $\Delta$, and the theory with the critical exponent $\beta$ as a \textsl{single} fit parameter. The data was logarithmically binned and averaged. No average was carried for detunings below $300$ Hz. Since we independently measured $\Gamma_0$ it is not a fit parameter. The measured signal exhibits a narrow resonance feature with a width of $\Gamma_0=45$ Hz, accompanied with long tails ranging up to $8$ kHz. 
We found $\beta=0.56 \pm 0.01$. We also reconstructed the absorptive part \cite{epaps} and observed clear deviations from a Lorentzian fit for the $1d$ beam as shown in \cite{RamseyDiffusion}.
 
 In Fig. \ref{fig:two}, we present our main result. In this graph, we plot the absolute value of the complex EIT spectrum derived from the measured dispersive part for the one-dimensional beam geometry (triangles) and for the wide area beam (circles) as a function of the complex detuning size $\abs{s}=\sqrt{\Delta^2+(\Gamma_0/2)^2}$ in units of $\Gamma_0/2$. 
In the universal regime, the complex spectrum is $S(s)=s^{-\beta}, \beta=0.5$. By writing $s$ in polar representation, $s=\sqrt{\Delta^2+(\Gamma_0/2)^2}e^{-i\arctan(2\Delta/\Gamma_0)}$, the dispersive part of the complex spectrum is $\abs{s}^{-\beta} \sin[\beta \arctan(2\Delta/\Gamma_0)]$.  In the experiment $\Delta$ was known and $\Gamma_0$ was measured. If we let $\beta$ to be a single fit parameter and search for the best fit for the measured data (see Fig. 1), we obtain in the $1d$ case $\beta=0.56 \pm 0.01$ over two decades in the frequency domain. For the wide area beam we obtain $0.97 \pm 0.01$ over two decades. After obtaining an estimate for $\beta$ we divide the measured spectrum by $\sin[\beta \arctan(2\Delta/\Gamma_0)]$ to obtain the absolute value of the complex spectrum [Fig. (\ref{fig:two})]. This should also be compared to the expected $\beta=0.52$ and $\beta'=0.99$  [Fig. (\ref{fig:three})].   
  
 In Fig. \ref{fig:three}, we plot the critical exponent as a function of the normalized diffusion broadening $\Gamma_{D}/\Gamma_0$ and the normalized power broadening $\Gamma_{p}/\Gamma_{0}$, which we calculated from the \emph{exact} solution of Eq. (5). We verified that our chosen beam size and power are within the asymptotic regime where the universal theory is correct. Marked with a triangle is the expected critical exponent ($0.52$) for the parameters used in our experiment. For the $2d$ wide-area beam the expected critical exponent based on the exact solution \cite{OferPRA} is $0.99$. In this case we verified that the presence of walls has a negligible effect on the critical exponent. We also studied the dependence on the optical depth (OD). For $OD\leq 10$ the effect on the critical exponent is less than $0.5$ percent. Higher values may alter the line-shape considerably due to the emergence of non-linear effects \cite{HigherOD}.
 
 The measured Laplace transform of the recurrence probability is in one-to-one correspondence with the FRT distribution in the time domain. In particular, when $\text{FRT}(s) \sim 1-s^{\beta}$, and $\beta<1$ as in our $1d$ case, the asymptote of $\text{FRT}(t)$ scales like $t^{-\beta-1}$ \cite{Redner}. It follows that \textit{the mean first return time is divergent} in the $1d$ case. This stands in stark difference to the probability to stay in the dark presented in \cite{RamseyDiffusion2}.  
  
  In conclusion, we found a regime where the EIT spectrum has a universal form, which is the Laplace transform of the recurrence probability that depends only on the effective dimensionality of the problem.   
Our result can be generalized to any kind of classical dynamics by replacing the diffusion propagator in Eq. (5) by the relevant center-of-mass propagator $G(r,r',t)$ e.g. of anomalous diffusion, or of ballistic motion in billiards. In that case, following the same derivation it is found that the EIT spectrum in the universal regime is of the form $G(0,0,s)=P(0,s)$.   
Staying in the realm of diffusion, our method can facilitate measurements of interesting FRT and FPT distributions in more complex settings such as diffusion in periodic and disordered (speckled) beam configurations and diffusion in the presence of traps. The latter are of particular interest, since the asymptotic survival probability of the dark state in their presence is anomalous and has the form of a stretched exponential \cite{Redner}. Simpler extensions are measuring the universal spectrum at $2d$ and measuring the Laplace transform of the first passage time from $\mathbf{r}=0$ to $\mathbf{r} \neq 0$, which is proportional to the coherence at this point $R_{21}(\mathbf{r},s)$. 
\bibliographystyle{apsrev}  

\begin{thebibliography}{17}
\bibitem[{rem()}]{remark}
\bibinfo{note}{The average is over all initial positions located within the
  beam. For $s$ close to zero $\text{FPT}(s) \sim 1/s$ \cite{OferPRA}.}
\bibitem[{rem()}]{zmotion}
\bibinfo{note}{The dark state change sign along the z-direction after $\lambda_z \sim 2.2$ cm. Prior to its decay a coherent atoms moves a charectaristic distance of $\sqrt{D/\Gamma_0} \sim 1 cm < \lambda_z$.}
\end{thebibliography}

\end{document}